\documentclass[12pt,preprint]{aastex}
\usepackage{textcomp}

\slugcomment{To appear in ApJL} \shorttitle{Mid-Infrared
Diagnostics of LINERs} \shortauthors{Sturm et al.}

\begin{document}
\title{Mid-Infrared Diagnostics of LINERs
}

\author{
E. Sturm, \altaffilmark{1}
D. Rupke, \altaffilmark{2}
A. Contursi, \altaffilmark{1}
D.-C. Kim, \altaffilmark{2}
D. Lutz,  \altaffilmark{1}
H. Netzer,  \altaffilmark{1,3}
S. Veilleux, \altaffilmark{2}
R. Genzel, \altaffilmark{1}
M. Lehnert, \altaffilmark{1}
L.~J. Tacconi, \altaffilmark{1}
D. Maoz, \altaffilmark{3}
J. Mazzarella, \altaffilmark{4}
S. Lord, \altaffilmark{4}
D. Sanders, \altaffilmark{5}
A. Sternberg \altaffilmark{3}
 }
\altaffiltext{1}{Max-Planck-Institut f\"ur extraterrestrische
Physik, Postfach 1312, D-85741 Garching, Germany;
sturm@mpe.mpg.de}
\altaffiltext{2}{Department of Astronomy, University of Maryland,
College Park, MD 20742, USA}
\altaffiltext{3}{School of Physics
and Astronomy, Tel Aviv University, Ramat Aviv, Tel Aviv 69978,
Israel}
\altaffiltext{4}{IPAC/Caltech, MS 100-22, Pasadena, CA
91125, USA}
\altaffiltext{5}{Institute for Astronomy, University, of Hawaii,
2680 Woodlawn Drive, Honolulu, HI 96822, USA}

\begin{abstract}
We report results from the first mid-infrared spectroscopic study
of a comprehensive sample of 33 LINERs, observed with the {\it
Spitzer Space Telescope}. We compare the properties of two
different LINER populations: infrared-faint LINERs, with LINER
emission arising mostly in compact nuclear regions, and
infrared-luminous LINERs, which often show spatially extended
(non-AGN) LINER emission. We show that these two populations can
be easily distinguished by their mid-infrared spectra in three
different ways: (i) their mid-IR spectral energy distributions
(SEDs), (ii) the emission features of polycyclic aromatic
hydrocarbons (PAHs), and (iii) various combinations of IR
fine-structure line ratios. IR-luminous LINERs show mid-IR SEDs
typical of starburst galaxies, while the mid-IR SEDs of IR-faint
LINERs are much bluer. PAH flux ratios are significantly different
in the two groups. Fine structure emission lines from highly
excited gas, such as [O\,IV], are detected in both populations,
suggesting the presence of an additional AGN also in a large
fraction of IR-bright LINERs, which contributes little to the
combined mid-IR light. The two LINER groups occupy different
regions of mid-infrared emission-line excitation diagrams. The
positions of the various LINER types in our diagnostic diagrams
provide important clues regarding the power source of each LINER
type. Most of these mid-infrared diagnostics can be applied at low
spectral resolution, making AGN- and starburst-excited LINERs
distinguishable also at high redshifts.
\end{abstract}


\section{Introduction}
\label{s:intro}

Since their identification as a class of galactic nuclei more than
25 years ago (Heckman 1980), the nature of Low-Ionization Nuclear
Emission-Line Regions (LINERs) has remained controversial.  Their
optical spectra are characterized by enhanced narrow emission
lines of low ionization species, quite distinct from those of both
\ion{H}{2} regions and classical active galactic nuclei (AGNs).
They are found in one third to one half of nearby galaxies of all
types (e.g., Ho, Filippenko, \& Sargent 1997). In many LINERs the
emission is concentrated near the nucleus (a few 100 pc, e.g.
Pogge et al. 2000) but in others it extends over larger regions,
up to a few kpc (Veilleux et al. 1995). There is substantial
evidence that many LINERs are powered by accretion onto massive
black holes, and that these objects, due to low accretion rates,
constitute the low-luminosity end of the AGN class (Quataert 2001,
Kewley et al. 2006). If many LINERs at low and high redshifts are
indeed low-luminosity AGNs, this
would have a significant impact on major issues in astronomy such
as the growth history of central black holes and the relation of
AGNs to galaxy formation and evolution.

Alternative scenarios for LINER excitation mechanisms have been
suggested. Models focusing on photoionization by the central AGN
continuum (e.g., Ferland and Netzer 1983; Groves et al. 2004),
have been complemented by stellar photoionization modeling (e.g.,
Barth and Shields 2000) and observations (e.g., Maoz et al. 1998)
showing that, under certain conditions, a young starburst can also
excite a LINER. In many cases LINER-like emission is also observed
on larger spatial scales; these `extended' LINERs tend to be
associated with high infrared luminosity (see
\textsection\ref{s:sample}). In addition to nuclear (Black Hole or
stellar) photoionization, these sources can be ionized by shock
heating through cloud collisions induced by accretion, galaxy
interactions, mergers, or starburst-driven winds (e.g. Shull \&
McKee 1979; Veilleux \& Osterbrock 1987; Dopita \& Sutherland
1995).

Determining which LINERs are photoionized by a hard, nuclear power
source and which are excited by other ionization processes is
crucial for understanding not only the local LINER population, but
also AGNs and starbursts  at high redshifts, if the fraction of
LINERs among high-$z$ galaxies is as high as in the local
universe. Galaxies with high infrared luminosities and galaxy
interactions are much more common at high redshifts than at
z$\approx$0 (e.g. P{\'e}rez-Gonz{\'a}lez et al. 2005). Whether or
not these galaxies host dominant AGNs is an open question. X-ray
and radio observations help to identify AGNs in the nuclei of
these galaxies. However, they may face problems in deriving the
{\it relative contributions} of AGN and starbursts to the
luminosity of those objects, in particular for close to
Compton-thick AGNs (e.g. NGC6240, Lutz et al. 2003). In order to
help resolve these problems we have performed a mid-infrared
spectroscopic study which is based on a large sample of
infrared-faint and infrared-luminous LINERs. This study is the
first of its kind, since only a few LINERs were observed with ISO
as part of various projects with differing scientific goals
(Satyapal, Sambruna, \& Dudik 2004), resulting in a sparse
collection of mid-IR LINER line detections of rather low S/N in a
heterogenous sample. Comprehensive mid-infrared spectroscopic
studies of LINERs have become possible only now with {\it
Spitzer}.

\section{Sample Selection, Observations and data Processing}
\label{s:sample}

We have based our selection of LINERs on the infrared luminosity
of low redshift objects. Most of the IR-faint LINERs
(L$_{IR}$/L$_B \lesssim 1$, with L$_{IR}$, the 8-1000$\mu$m
luminosity, defined as in Sanders \& Mirabel 1996) studied in the
literature have compact nuclei (e.g. Pogge et al. 2000),
consistent with photoionization from a central source. This
central source is very likely an AGN, since most of these IR-faint
LINERs have compact nuclear X-ray sources, but no extended X-ray
sources (e.g. Carrillo et al. 1999, Satyapal et al. 2004). On the
other hand, IR-luminous LINERs (L$_{IR}$/L$_B \gtrsim 1$) often
also show compact hard X-ray cores indicative of an AGN, but in
addition they often contain multiple off-nuclear X-ray point
sources. There is evidence for extended LINER emission in a number
of LINERs in the Revised Bright Galaxy Sample (RBGS, see Veilleux
et al. 1995). A famous example is the extended LINER emission in
NGC 6240 (Veilleux et al. 2003). Monreal-Ibero et al. (2006) also
show IFU [NII]/H$\alpha$ excitation maps of IR-luminous LINERs
showing extended LINER emission. Our two sub-samples of different
IR-luminosity therefore represent statistically different LINER
populations (nuclear versus extended LINER emission).

The sample is summarized in Table \ref{tbl:sample}. We selected 16
IR-luminous LINERs from the RBGS, and 17 IR-faint LINERs from Ho
et al. (1997). The IR-faint LINERs are further subdivided into 5
Type\,1 and 6 Type\,2 LINERs, and in addition 6 ``transition''
objects (see Ho et al. 1997, and \textsection\ref{s:results_SED}).
The infrared-bright LINERs are all of Type\,2. With a median
L$_{IR}$ of 11.31 L$_\odot$ the IR-luminous LINERs are much more
luminous than the IR-faint LINERs in the IR, and there is no
significant difference between the IR-faint sub-classes (L1: 8.92
L$_\odot$, L2: 9.27 L$_\odot$, T2: 9.31 L$_\odot$). For further
multi-wavelength information for all our objects we refer to
Carillo et al. (1999). Table \ref{tbl:sample} also lists
L$_{IR}$/L$_{B}$ ratios. The median ratios are 106 and 1.8 for the
IR-bright and IR-faint LINERs respectively.
The data were obtained with the InfraRed Spectrograph (IRS; Houck
et al. 2004) on board {\it Spitzer} in staring mode. High
resolution spectra covering $10-37~\micron$ (modules SH and LH)
are complemented by low-resolution spectra in the $5-15~\micron$
range (modules SL2 and SL1). Our data reduction process started
with the two dimensional BCD products from the Spitzer pipeline
(version S12). We used our own IDL routines to perform
de-glitching, and SMART (Higdon et al. 2004) for the final
spectrum extraction.

All our targets are nearby (D$\le$ 120Mpc), but on average the
IR-faint LINERs have lower redshifts (median z=0.004) than the
IR-bright objects (median z=0.023). We carefully checked that
aperture effects (extended objects at different distances, and
fluxes from different aperture sizes in the different IRS modules)
do not significantly affect the spectra and our conclusions. For
most of the targets the 12 and 25 $\mu$m flux densities agree
quite well with the IRAS values (the worst deviation is a factor 2
in NGC5371, a nearby, IR-faint LINER which is dominated by mid-IR
emission from the spiral arms); flux density discrepancies in the
overlapping wavelength region between sub-spectra obtained with
different aperture sizes are of the same order; and the profiles
of most of the two-dimensional spectral images do not deviate
significantly from stellar (point source) profiles. I.e. the
observed flux in most of the objects is coming from a nuclear
region of 0.3 kpc (median) diameter. We did not apply an extended
source correction to the flux calibration, and we used full width
apertures for the spectral extraction. More details about the
sample, the observations and the data processing will be described
in a forthcoming paper (D. Rupke et al., in prep.).

\section{Mid-infrared properties of LINERs}
\label{s:results}

Our new {\it Spitzer} observations provide three ways to
distinguish among the various sub-classes of LINERs: the
mid-infrared SEDs, the PAH emission features, and various
combinations of IR fine-structure line ratios. These are
illustrated and discussed below. Figure \ref{F:average_spectra}
shows the average spectra of both sub-samples (with the IR-faint
LINERs further divided into the three subgroups Type\,1, Type\,2,
and transition objects, see \textsection\ref{s:sample}). As
described in \textsection\ref{s:sample} the IR-faint LINERs,
including the transition objects, are much less luminous in the IR
than the IR-luminous ones. Before averaging, the individual
spectra have been normalized at 19$\mu$m. The average spectra show
clear differences among the sub-groups. Because of a low
dispersion within the two major groups (IR-luminous, IR-faint)
almost all of our individual spectra can be unambiguously assigned
to one of these two groups (Figure \ref{F:shocking_diagram}).

\subsection{SED shape}
\label{s:results_SED}

\begin{figure*}[t]
\plotone{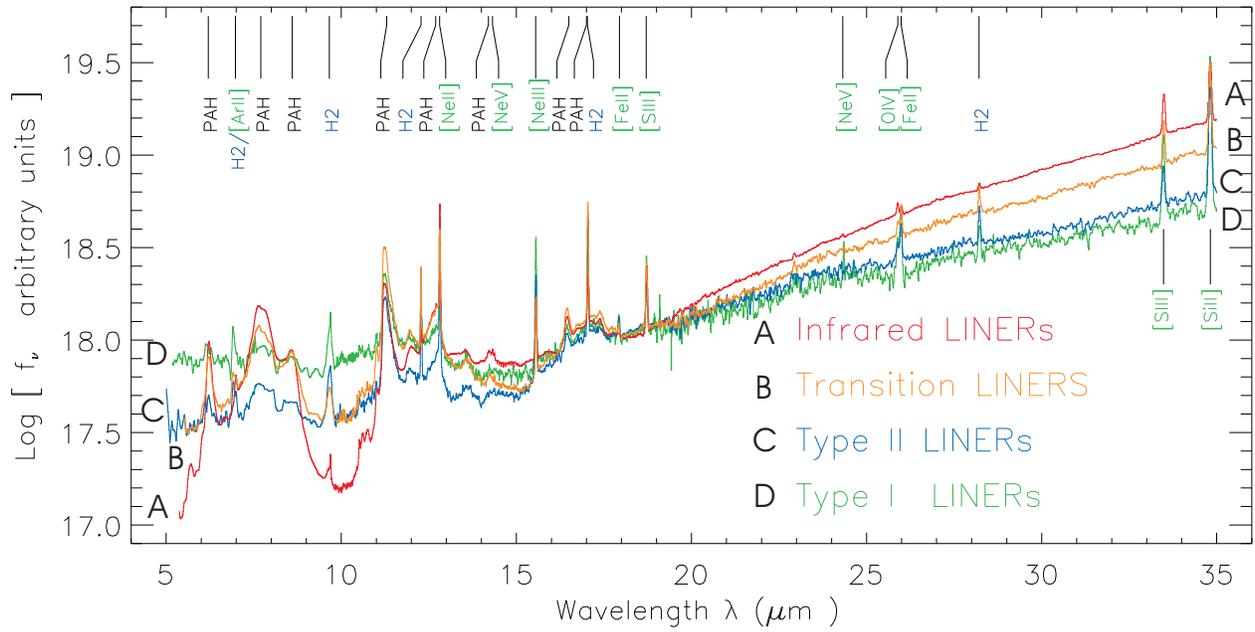} \caption{\footnotesize Composite spectra
(normalized at 19$\mu$m) of the 16 IR-luminous LINERs (red) and
the IR-faint LINERs (green: 5 Type\,1 LINERs, blue: 6 Type\,2
LINERs, yellow: 6 Transition Type).} \label{F:average_spectra}
\end{figure*}

\begin{figure*}[t]
\plotone{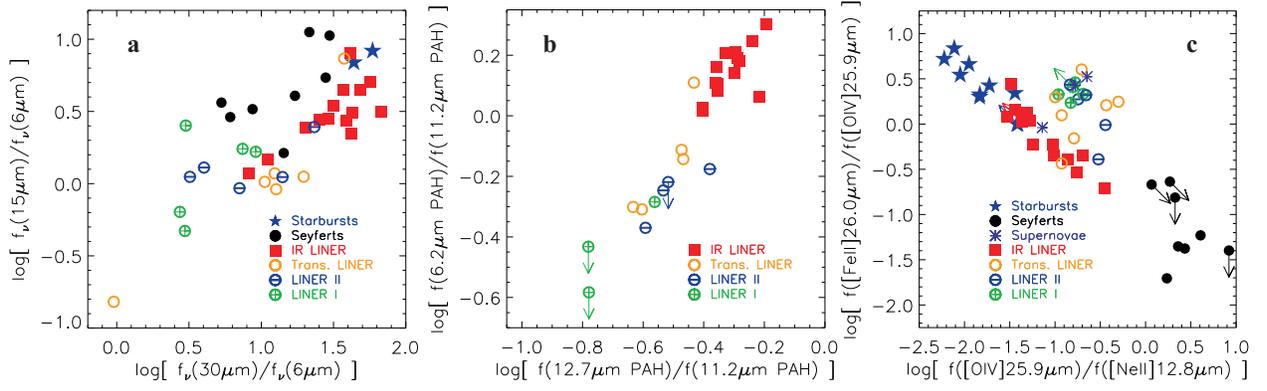} \caption{\footnotesize Mid-infrared diagnostic
diagrams to distinguish between IR-luminous and IR-faint LINERs.
In panel {\it a} Seyfert data are from Weedman et al. (2005),
Starbursts from Sturm et al. (2000, M82) and Brandl et al. (2004,
NGC7714); Starburst and Seyfert data in panel {\it c} are from
Sturm et al. (2002) and Verma et al. (2003). Error bars in all
three panels are comparable to the symbol sizes.}
\label{F:shocking_diagram}
\end{figure*}

The mid-infrared SEDs of IR-faint LINERs are clearly flatter
(i.e., bluer) than their IR-luminous counterparts. No strong
absorption features due to silicates (at 9.7 and 18 $\micron$) or
ices (at various wavelengths) are seen in either sub-sample.  Only
one IR-faint LINER, NGC~3998, reveals strong silicate emission
(Sturm et al. 2005). In contrast, the mid-infrared SEDs of
infrared bright LINERs generally look very similar to those of
starburst galaxies (e.g., Sturm et al. 2000). There are
interesting trends in Figure \ref{F:average_spectra}. The Type\,1
and Type\,2 spectra are very similar, but below $\sim$15$\mu$m the
average Type\,1 spectrum has an additional warm (dust) component.
The average Transition LINER lies in between the Type\,1/Type\,2
objects and the IR-bright LINERs. These transition objects are
classified as objects with [\ion{O}{1}] strengths intermediate
between those of \ion{H}{2} nuclei and LINERs, which could be
explained either by a dilution of ``true'' LINER spectra with
circumnuclear star formation or by stellar photoionization (see Ho
et al. 1997).

The differences in the mid-IR SEDs of the two major LINER types
are quantified in Figure \ref{F:shocking_diagram} (left panel): in
a diagram of the continuum flux ratios
f$_\nu$(15$\mu$m)/f$_\nu$(6$\mu$m) vs.
f$_\nu$(30$\mu$m)/f$_\nu$(6$\mu$m) IR-faint and IR-bright galaxies
are nicely separated. The warmer spectrum of (some of) the Type\,1
LINERs compared to the Type\,2 LINERs (see also Fig.
\ref{F:average_spectra}) could be the signature of a warm AGN
continuum (as expected in a comparison of Type\,1 and Type\,2
AGNs, assuming the AGN unification model holds for LINERs). Maoz
et al. (2005) have found a similar trend in the UV continuum of
IR-faint LINERs. In addition, in some of the weaker, nearby
IR-faint LINERS photospheric emission from old stellar populations
in the center of the host galaxies may contribute significantly to
the measured continuum in the range 5 to
$\sim$8 $\mu$m. Using the same scaling factor as Lutz et al.
(2004) to extrapolate from K-band to mid-IR, and assuming that
most of the K-band flux in our IRS aperture is stellar, we
conclude that this might indeed be the case for some of the
IR-faint LINERs. Stellar contribution is probably not the sole
explanation, however, because it could hardly explain the
difference between (the average) Type\,1 and Type\,2 LINERs (the
K-band/15$\mu$m ratios are similar in both types), and because the
typical signature of such a component, a decreasing continuum in
the 5-8$\mu$m range, is not prominent in most of the individual
spectra.

\subsection{PAH ratios}
Figure \ref{F:average_spectra} also shows significant differences
in the PAH features in the spectra of the two LINER groups. In
IR-bright LINERs the PAH spectrum is very similar to starburst
galaxies. IR-faint LINERs have very weak PAH features in the
$5-10~\micron$ range, but a strikingly strong 11.2 $\mu$m feature.
We note that, in principle, a higher degree of dilution by an
additional warm dust component in the IR faint LINERs could cause
errors in the flux measurements of PAHs in the $5-10~\micron$
range, if the continuum has structure of similar width as the
PAHs. For instance, strong silicate emission or absorption around
10 $\mu$m, if not accounted for, could lead to systematically
false flux values of broad PAH features like the 7.7/8.6 $\mu$m
PAH complex. Such an effect, however, should not play a
significant role in our measurements of the relatively narrow,
well defined PAHs at 6.2, 11.2 and 12.7 $\mu$m. Based on the flux
ratios of these PAH features at 6.2, 11.2, and 12.7 $\mu$m, we
present in Figure \ref{F:shocking_diagram} (middle panel) another
diagnostic diagram for the distinction between IR-bright and
IR-faint LINERs. Hony et al. (2001, their Figure 5) have
constructed such a diagram using ISO spectra of galactic HII
regions, YSOs and evolved stars/reflection nebulae. They found
that the strongest 11.2$\mu$m PAH features (lowest 12.7/11.2 and
6.2/11.2 ratios) are produced in evolved stars, while HII regions
have the highest 12.7/11.2 (and 6.2/11.2) ratios. They attributed
this to a different degree of ionization (the 11.2$\mu$m PAH
carriers are neutral, those of the 6.2 and 12.7 features are
ionized; see also Joblin et al. 1996), as well as a processing of
PAHs from large (11.2$\mu$m) to smaller (12.7$\mu$m) carriers.
This could imply a less ionizing environment in the IR-faint
LINERs than in the IR-bright ones, or significant differences in
PAH formation and destruction.

\subsection{Emission line ratios}

The mid-infrared LINER spectra are rich in fine structure emission
lines. We have detected strong [O\,IV] lines, i.e. highly ionized
gas, in both LINER sub-samples, suggestive of the presence of an
AGN in $\sim$90\% of the LINERs. The presence of a weak AGN in
many of the IR-luminous LINERs is consistent with the compact hard
X-ray cores that many IR-bright LINERs show (see
\textsection\ref{s:sample}), and is further supported by the
detection of [Ne\,V] in about half of these sources. This
detection rate is certainly a lower limit due to the limited S/N
in the [Ne\,V] spectra. In order to explore differences in
relative line strengths and physical properties, we have
constructed mid-infrared diagnostic diagrams involving various
combinations of these fine structure emission lines. The right
panel in Figure \ref{F:shocking_diagram} is an example of such a
diagram using the line ratios of
[\ion{Fe}{2}]26.0$\mu$m/[\ion{O}{4}]25.9$\mu$m and
[\ion{O}{4}]25.9$\mu$m/[\ion{Ne}{2}]12.8$\mu$m. Lutz et al. (2003)
used this diagram to distinguish among excitation by early type
stars, AGNs, and shocks. Starburst galaxies, Seyfert galaxies, and
supernova remnants (SNRs) are clearly separated in this line ratio
plane. Interestingly, IR-luminous and IR-faint LINERs are easily
distinguishable in this diagram: the IR-luminous LINERs all lie on
a linear relation connecting starbursts and Seyferts, which might
be explained by a minor AGN contribution in addition to the star
forming regions (see mixing lines in Lutz et al. 2003). The
position of the IR-luminous LINERs in this diagram is an
indication, that the weak AGN in these objects is more
Seyfert-like, i.e. not responsible for the IR-faint LINER
emission. Surprisingly, the IR-faint LINERs are clearly off this
line and lie in a region populated by SNRs. The reason for the
differences between the two classes (and between IR-faint LINERS
and Seyferts) is so far uncertain. The role of the hardness of the
radiation spectrum and of the ionization parameter, as well as the
influence of shocks in both LINER classes has to be further
examined. Can, e.g., photoionized spectra for certain hardness and
ionization parameters reach the `SNR' region in Figure 2c?

\section{Conclusions and Outlook}

Our systematic study of LINERs with {\it Spitzer} confirms that
IR-luminous sources of this group are very different in their
properties from IR-faint LINERs. IR-luminous LINERs have infrared
SEDs that are similar to those of starburst galaxies, and they are
situated in different regions of several diagnostic diagrams than
IR-faint LINERs. While this has been suspected from previous
multi-wavelength studies, the new mid-IR data provide the cleanest
way to separate those groups. Many of these differences are also
visible at low spectral resolution, and thus at high redshift,
making our discovery an important cosmological tool for discerning
the role of AGN in galaxy formation and evolution. A more detailed
investigation of this issue and a quantification of the relative
contributions from AGN and star formation to the LINER spectra
requires a careful disentangling of the spectra into the various
components (stellar, HII region, AGN, etc.), involving template
fitting methods. Furthermore, for an understanding of the emission
line properties, detailed photoionization modeling is required.
This is the subject of a forthcoming paper (D. Rupke et al., in
preparation).

\acknowledgments

This work is based on observations made with the Spitzer Space
Telescope, which is operated by the Jet Propulsion Laboratory,
California Institute of Technology under a contract with NASA.
Support for this work was provided by NASA through an award issued
by JPL/Caltech.
AS and HN thank the
Israel Science Foundation for support grants 221/03 and 232/03. HN
acknowledges support by the Humboldt Foundation and thanks the
host institution MPE.

%
\vspace*{0.3cm}
\newpage
\noindent {\bf  \large References}\\[0.2cm]
Barth, A.J. \& Shields, J.C. 2000, PASP, 112, 753\\
Brandl, B. et al., 2004, \apjs, 154, 188 \\
Bressan et al. 1998,  \aap, 332, 135 \\
Bressan et al. 2006, \apjl, 639, L55 \\
Carrillo, R., et al. 1999, RMxAA, 35, 187\\
Dopita, M.~A., \& Sutherland, R.~S.\ 1995, \apj, 455, 468 \\
Ferland, G.~J., \& Netzer, H.\ 1983, \apj, 264, 105 \\
Groves et al.\ 2004, \apjs, 153, 9 \\
Heckman. T.M. 1980, A\&A, 87, 152\\
Ho, L.C., et al. 1997, ApJS, 112, 315\\
Hony, S., et al.\ 2001, A\&A, 370, 1030\\
Higdon, S.~J.~U., et al.\ 2004, \pasp, 116, 975 \\
Houck, J.~R., et al.\ 2004, \apjs, 154, 18 \\
Joblin C., Tielens A.G.G.M., Geballe T.R., Wooden D.H. 1996, ApJ, 460, L119\\
Kewley, L.~J. et al. 2006, astro-ph/0605681\\
Lutz, D., et al. 2003,A\&A, 409, 867\\
Lutz, D., et al. 2004,A\&A, 418, 465\\
Maoz, D. et al. 1998, AJ, 116, 55\\
Maoz, D., Nagar, N.~M., Falcke, H., \& Wilson, A.~S.\ 2005, \apj,
625, 699 \\
Monreal-Ibero, A., Arribas, S., \& Colina, L.\ 2006, \apj, 637,
138\\
Monreal-Ibero, A., Arribas, S., \& Colina, L.\ 2006, astro-ph/0509681\\
P{\'e}rez-Gonz{\'a}lez, P.~G., et al.\ 2005, \apj, 630, 82\\
Pogge, R.~W. et al. 2000, ApJ, 532, 323\\
Quataert, E. 2001, ASP Conf. Proc., 224, p.71\\
Sanders, D.~B., \& Mirabel, I.~F.\ 1996, \araa, 34, 749\\
Satyapal, S., Sambruna, R.~M., \& Dudik, R.~P.\ 2004, \aap, 414,
825\\
Shull, J.~M., \& McKee, C.~F.\ 1979, \apj, 227, 131 \\
Sturm, E., et al. 2000, A\&A, 358, 481\\
Sturm, E., et al. 2005, A\&A, 629, L21\\
Veilleux, S., et al.  1995, ApJS 98, 171\\
Veilleux, S. \& Osterbrock,  1987, ApJS, 63, 295\\
Verma, A., et al. 2003, A\&A, 403, 829\\

\clearpage

\begin{deluxetable}{llllc|llllc}
\tabletypesize{\scriptsize}
\tablecaption{The LINER Sample
\label{tbl:sample}}
\tablewidth{0pt}
\tablehead{ \colhead{Object Name} & \colhead{Catalog} &
\colhead{Type} & \colhead{L$_{IR}$\tablenotemark{a}} &
\colhead{L$_{IR}$/L$_B$\tablenotemark{b}} & \colhead{Object Name}
& \colhead{Catalog} & \colhead{Type} &
\colhead{L$_{IR}$\tablenotemark{a}} &
\colhead{L$_{IR}$/L$_B$\tablenotemark{b}}}
\startdata
UGC556      & RBGS\tablenotemark{c}& L2 &10.91 & 106 & NGC404  & Ho97\tablenotemark{d} & L2 & 7.83 & 1.1\\
UGC2238     & RBGS& L2  & 11.35 & 93 & NGC3245 & Ho97 & T2 & 9.48 & 1.8\\
IRAS02438+2122& RBGS& L2& 11.20 & 205& NGC3507 & Ho97 & L2 & -    & -  \\
NGC1204     & RBGS& L2  & 10.96 & 113& NGC3642 & Ho97 & L1 & 9.65 & 1.4\\
IRAS05187-1017&RBGS& L2 & 11.31 & 229& NGC3884 & Ho97 & L1 & -    & -  \\
NGC4666     & RBGS& L2  & 10.44 & 29 & NGC3898 & Ho97 & T2 & 9.13 & 0.6\\
NGC4922     & RBGS& L2  & 11.31 & 103& NGC3998 & Ho97 & L1 & 8.66 & 0.6\\
UGC8387     & RBGS& L2  & 11.65 & 135& NGC4036 & Ho97 & L1 & 9.18 & 0.5\\
NGC5104     & RBGS& L2  & 11.19 & 79 & NGC4192 & Ho97 & T2 & 9.72 & 4.7\\
NGC5218     & RBGS& L2  & 10.64 & 21 & NGC4278 & Ho97 & L1 & 8.50 & 0.5\\
IZw107      & RBGS& L2  & 11.89 & -  & NGC4419 & Ho97 & T2 & 9.91 & 11 \\
IRAS15335-0513& RBGS& L2& 11.42 & -  & NGC4435 & Ho97 & T2/H & 9.09 & 1.9\\
IRAS16164-0746& RBGS& L2& 11.44 & 210& NGC4457 & Ho97 & L2 & 9.27 & 4.3\\
ESO602-G025 & RBGS& L2  & 11.33 & 47 & NGC4486 & Ho97 & L2 & 8.95 & 0.1\\
Zw453.062   & RBGS& L2  & 11.38 & 116& NGC5371 & Ho97 & L2 & 10.59\tablenotemark{e} & 4.9\\
NGC7591     & RBGS& L2  & 11.12 & 47 & NGC6500 & Ho97 & L2 & 9.74 & 3.0\\
            &     &     &       &    & NGC7177 & Ho97 & T2 & -    & -  \\

\enddata
\tablenotetext{a}{Logarithm of the 8-1000$\mu$m luminosity in
units of the solar luminosity}%
\tablenotetext{b}{L$_B$ from B magnitudes in Carillo et al. (1999)}%
\tablenotetext{c}{RBGS= Revised Bright Galaxy Survey, Veilleux et
al. (1995)}%
\tablenotetext{d}{Ho97=Ho et al. (1997)} %
\tablenotetext{e}{nuclear LINER from Ho97, with much of the mid-IR
luminosity coming from spiral arms outside the IRS apertures,
hence treated as IR-faint LINER}
\end{deluxetable}

\clearpage

\clearpage

\end{document}